# Atmospheric rivers and winter sea ice drive recent reversal in Antarctic ice mass loss

Marlen Kolbe[1*], Jose Abraham Torres Alavez[2], Ruth Mottram[2], Marwan Katurji[3], Richard Bintanja[4,5], Eveline C. van der Linden[5]

[1*]Department of Earth, Ocean and Atmospheric Sciences, University of British Columbia, Vancouver, BC, Canada.
[2]National Centre for Climate Research, Danish Meteorological Institute, Copenhagen, Denmark.
[3]Faculty of Science, University of Canterbury, Christchurch, New Zealand.
[4]Faculty of Science and Engineering, Energy and Sustainability Research Institute Groningen, University of Groningen, Groningen, Netherlands.
[5]Research and Development of Weather and Climate models, Royal Dutch Meteorological Institute, De Bilt, Netherlands.

*Corresponding author. E-mail: mkolbe@eoas.ubc.ca;

**Abstract**

Since about 2000, the total mass of the Antarctic Ice Sheet (AIS) has declined at a near-linear rate, increasing global sea levels. Since 2016 however, satellite gravimetry data reveal a slowdown in net AIS mass loss and a net mass gain since 2020, despite increases in dynamically-driven ice loss by discharge from outlet glaciers. Here we use a suite of reanalyses and regional climate models to show that this reversal is caused by increased precipitation and positive surface mass balance anomalies linked to increased atmospheric river (AR) activity, strengthening westerlies and loss of sea ice. ARs have become more frequent and intense since 2020, particularly over the Antarctic Peninsula, Queen Maud Land, and Wilkes Land, resulting in strong regional positive mass balance anomalies. High-resolution regional climate model simulations with modified sea ice extent show that the effect of sea ice on enhancing precipitation through increased evaporation accounts for around 10% of the winter increase, but is overall minor compared to remote large-scale processes. Combined, these factors result in accumulation increases that currently offset the mass loss from accelerated ice discharge in Antarctica and point to processes important for future projections.

Understanding the drivers of mass gains and losses for the Antarctic Ice Sheet (AIS) is crucial for predicting future sea level rise and changes in ocean circulation. Antarctica holds the largest reservoir of freshwater on Earth, and therefore changes in its ice mass balance strongly influences global and regional sea levels. Related freshwater influx and changes in the sea ice extent (SIE) can further affect density-driven global ocean circulation [1–4]. AIS mass balance has shown a near-linear decline since 2002, but recent studies based on gravimetry-based estimates from the GRACE satellite missions (GRACE and its successor, GRACE-FO) reveal an abrupt slowdown in mass loss between 2020 and 2022 [5, 6]. This interruption in the downward trend has been attributed to increased precipitation and surface mass balance (SMB) anomalies, which are unprecedented in both West and East Antarctica [5, 7]. Updated GRACE data through December 2024 shows that ice mass accumulation has accelerated further due to even greater precipitation since 2022. A new ice discharge dataset [8] confirms that this trend is not due to reduced ice discharge (which continues to accelerate), but is instead driven by enhanced surface accumulation, which more than compensates for the enhanced ice loss in the total mass budget. On the AIS, extreme precipitation events such as atmospheric rivers (ARs), are the primary source of precipitation [9–12], which are highly episodic. In this study we show that their intensity and persistence have increased since 2020, particularly in areas where mass balance anomalies identified by the GRACE satellite data are highest, namely the Antarctic Peninsula (AP), Queen Maud Land and Wilkes Land. We examine the role of hemispheric and regional scale climate indices ENSO (El Nino Southern Oscillation) and the related Southern Annular Mode (SAM) in driving regional scale patterns. Additionally, we analyse the role of the low SIE in recent years by conducting pan-Antarctic experiments at 11km resolution in a regional climate model optimised for Antarctica with modified SIE for one year (July 2021 to June 2022), which includes four AR events, including the heatwave events of February 2022 on the AP [13, 14] and of March 2022 on the EAIS [15–18].

## Recent Increase in Antarctic Ice Mass

From the beginning of this century until 2020, Antarctic mass loss was estimated to range between -90 and -142 Gt yr$^{-1}$ [5, 6, 19–21]. After the launch of GRACE-FO in 2018, Zhang et al. (2024) reported a reduction in mass loss to only -24.8 $\pm$ 52.1 Gt yr$^{-1}$ until 2022 [6], and Wang et al. (2023) reported a mass gain of 129.7 $\pm$ 69.6 Gt yr$^{-1}$ from 2021 to 2022 [5]. Using GMB data to December 2024, we show that the recent years of mass gain were not short-term anomalies, but the beginning of a significant 5-year mass gain trend of 67.53 $\pm$ 31.4 Gt yr$^{-1}$ from 2020 to 2024 (Figure 1a). This positive trend occurs despite higher rates of dynamic ice loss, with the AIS losing nearly 100 Gt yr$^{-1}$ more ice through grounding line discharge between 2020–2024 compared to 2003–2019 (Figure 1d,g). Instead, the recent mass gain is driven by an increase in SMB (the sum of precipitation, evaporation and sublimation, and surface runoff), which remained near equilibrium until 2020 but has since then been rising at rates of ∼200 Gt yr$^{-1}$ (219.9 ± 14.9 Gt yr$^{-1}$ in ERA5; 197.42 ± 17.7 Gt yr$^{-1}$ in RACMO2.4p; and 196.37 ± 13.6 Gt yr$^{-1}$ in HCLIM43; Figure 1a). This increase represents ∼9% of the recently reported multimodel ensemble mean SMB of 2,300 Gt yr$^{-1}$ and exceeds the standard deviation of 108 Gt yr$^{-1}$ [22]. We show that most of the additional precipitation occurred over the ice shelves, where the increase in SMB is more pronounced than over grounded land (Figure A1). This suggests that ice shelves act as precipitation sinks or 'buffer zones' for moisture transported from further north, with the majority of precipitation reaching the AIS itself being primarily (up to 90%) driven by ARs [12]. The frequency of ARs reaching the AP and coastal East Antarctica is indeed significantly higher in the later period (2020–2024) compared to 2003–2019 (Figure A5), and coincides with the regions showing the strongest positive SMB and GMB trends (Figure 1e,f). Since 2020, only the drainage basins near the Amundsen Sea (Zwally drainage basins 19-21) still have a negative net mass loss due to high rates of dynamical discharge (Figure 1e,g). Annual precipitation over the AIS is dominated by short-term, high-impact events (e.g. as shown for 2021/2022 in Figure 2a), and recent precipitation increases are evident across most days of the year, particularly during summer (Figure 2b). This results in a 12.7% higher summer SMB in ERA5 (10% in RACMO2 and 12.2% in HCLIM43) during 2020–2024 compared to the 2003–2019 mean, while winter SMB increased by 7.3% in ERA5 (8.4% in RACMO2 and 5.8% in HCLIM43) (Figure A4). The summer increase is most pronounced over West Antarctica and the AP, whereas winter SMB has risen mainly along the East Antarctic coastline and tip of the AP. These patterns coincide with regions experiencing the highest AR activity (Figures A5 and A6) and IVT (Figure A7), as well as strengthened westerlies over the past five years (Figure A8). Given the relatively short duration (5 years) of increased SMB and AR activity, it remains unclear whether the recent GMB shift marks the onset of a long-term trend. As of April 2025, SMB has not shown signs of declining (Figure 1a), and continued increases in precipitation are expected under warming, consistent with the Clausius-Clapeyron relation [23].

## Regional Accumulation Patterns shaped by SAM, ENSO, and ARs

Previous studies have linked regional variation in Antarctic SMB to different phases of the SAM and the El Niño-Southern Oscillation (ENSO) [24–29]. Our correlation results align with the finding that the influence of SAM and ENSO on SMB and sea ice is complex and regionally dependent (Figure A2 and A3). Two dominant patterns emerge: (1) Both modes exert relatively weak influence on East Antarctica and the sea ice in the surrounding ocean basins; and (2) a positive SAM and negative ENSO tend to increase SMB on the AP and reduce it in the Ross Sea sector in both seasons (Figure A3), consistent with previous studies [24, 25, 28, 30]. The concurrent effect of SAM and ENSO on both sea ice and SMB near the AP is also described in an ice core study [26], which concluded that the anticorrelation between sea ice and SMB is likely not directly driven by sea ice loss, but by their shared response to large-scale atmospheric forcing. The recent sea ice decrease in the Bellingshausen Sea and the SMB increase on the AP since 2020 are thus very likely enhanced by negative ENSO and positive SAM anomalies respectively (Figure A3). This is also supported by the post-2020 alignment of SAM anomalies with Pan-Antarctic time series of SMB (Figure A2). The increased westerlies and deepened Amundsen Sea Low associated with a positive SAM and negative ENSO [25] have favoured more frequent intrusions of north-westerly air masses into the AP year-round, while reducing ARs, IVT and SMB over West Antarctica in winter (Figures A2–A3). However, SMB and GMB trends over East Antarctica since 2020 are just as strong (Figure 1e,f), and thus SAM and ENSO variability alone can not fully explain the enhanced accumulation. Previous studies have shown that East Antarctic precipitation is more strongly influenced by synoptic-scale moisture transport from northerly sources, largely due to the higher elevation of the plateau [10, 31]. This explains low correlations of inland East Antarctic SMB with local sea ice (Figure A10) and either climate modes (Figure A3a-f), and suggests that recent episodic AR events have instead increased precipitation in East Antarctica, especially during winter (Figure A5d and Figure A6d).

## The role of sea ice in modulating SMB and AR strength

### Correlations

Local sea ice loss can contribute to higher precipitation and SMB anomalies over the AIS [32, 33] by enhancing sensible and latent heat fluxes to the atmosphere [34, 35]. We find significantly negative correlations between seasonal SIC and SMB, especially during winter and in West Antarctic sectors (Figures A9 and A10). However, in some regions such as the Bellingshausen Sea this relationship partly reflects a shared response to large-scale atmospheric forcing [26]. Correlations between evaporation and adjacent SMB anomalies are generally weak or not statistically significant across most regions and seasons (not shown), but we note that ERA5-based evaporation estimates may be limited in accurately capturing latent heat fluxes due to simplifications in surface energy partitioning and parameterizations. We find that SIC correlations with SMB are generally stronger over ice shelves (Figure A10), which indicates that increased moisture from local sea ice loss is lost through precipitation over the ice shelves before reaching the grounded AIS. We also investigated different time lags (up to 3 months), as well as monthly and annual frequencies, which showed weaker correlations than the seasonal results presented here. Further, Antarctic sea ice began to decline since 2015, several years before the observed shifts in SMB and GMB trends (Figure 1a). The direct effect of sea ice loss on Antarctic mass gain through locally enhanced evaporation is therefore not insignificant, but unlikely to explain the sudden acceleration in ice sheet-wide precipitation.

### Idealised Experiments

To isolate the influence of sea ice on overall Antarctic-wide SMB as well as AR strength, we conducted 1-year idealized simulations from July 2021 to June 2022 with altered SIC (Figure 3a-c). Figure 2a displays the daily precipitation over the full year and highlights how episodic AR events account for a major share of the total accumulation. All four of the highest precipitation days (>15 Gt day$^{-1}$) over the AIS during this 1-year period occurred during ARs (Figure 4a-d).

The conducted RCM simulations with HCLIM43 [18, 36] include a control run (CTRL) and two idealised experiments with added sea ice (exICE, where SIC are increased to 100% up to 5°S north of the monthly mean SIE), and completely removed sea ice (noICE). The mean SIC of the different experiments over the simulated year are shown in Figure A11a-c. We note a 249 Gt increase in accumulated grounded AIS precipitation, and a 174 Gt increase in accumulated SMB from July 2021 to June 2022 when comparing noICE to CTRL (Figure 3d, Figure A11f). Our experiments thus suggest that a completely sea ice-free Southern Ocean during the already low sea ice in 2021/2022 would thus have increased annual grounded ice sheet precipitation by 8.8% and SMB by 7.3% (10.2% and 7% including ice shelves). Directly comparing the two extreme scenarios (exICE and noICE) results in slightly higher values, but here we focus on the changes relative to CTRL. Most additional precipitation from CTRL to noICE falls on ice shelves and coastal areas, where local precipitation is at some locations increased by as much as 1,000 mm year$^{-1}$ (Figure 3f). The areas most affected are Marie Byrd Land and the AP, with over 1,700 mm year$^{-1}$ of increased precipitation near Larsen C (Figure 3f). Correspondingly, precipitation is most reduced in these areas if SIC are increased (Figure 3e).

We also find higher temperatures over the AIS and ice shelves if sea ice is absent, particularly over the Antarctic Peninsula and near the Ross Ice Shelf (Figure A11g–i), where annual mean skin temperatures increase by up to 16°C. This warming explains why the SMB increase under removed sea ice is smaller than the corresponding increase in precipitation (Figure A11a-c), as a substantially larger portion of land and ice shelf becomes susceptible to evaporation (and runoff which is not included in our simplified ERA5 and HCLIM43 SMB calculation). The higher land and ice shelf temperatures in the noSIC experiment occur despite lower net downward longwave radiation and turbulent heat fluxes compared to the exICE and CTRL simulations (Figure A13-A15). Downward longwave radiation over both ocean and land is increased in noSIC, but this is outweighed by stronger upward longwave radiation, resulting in a net surface cooling effect from longwave radiation (Figure A13). Instead, higher annual mean land temperatures result primarily from increased net solar radiation (Figure A14) due to lower albedo (Figure A12d-f). Despite increased cloud water content (Figure A12a-c), which lowers downward shortwave absorption, future sea ice loss may therefore increase Antarctic temperatures primarily through reduced shortwave reflection, rather than turbulent fluxes or longwave radiation (though local exceptions exist). We also note a 15.4% increase in land rainfall in noICE (24% decrease in exICE), which is most evident on the AP and Victoria land (Figure A11d-f). Zonal and meridional winds are not significantly affected by the absence or presence of sea ice, suggesting that the recent strengthening of the westerlies (Figure A8) is instead driven by remote, planetary-scale atmospheric forcing.

These atmospheric changes in response to altered SIC are most pronounced during winter (Figure A16g–v), while the impact is minimal in summer (except for shortwave radiation), when a weaker ocean–air temperature gradient limits evaporation. Since recent mass loss and sea ice loss occurs in both summer and winter, this is further evidence that sea ice loss alone can not explain recent increases in summer SMB. We note a near-linear SMB increase from the exICE over the CTRL to the noICE experiment, with 5.7 (7.3) Gt more SMB per year per million km$^2$ of less SIE in summer (winter) over the AIS (Figure A17c). A backward calculation based on SMB and sea ice before and after 2020 suggests that the recent grounded ice sheet SMB increase of 121.9 Gt year$^{-1}$ in summer and 78.5 Gt year$^{-1}$ in winter (Figure A4) can be attributed by approximately 3.1% and 10.9% to sea ice loss, respectively. This is based on the fact that the mean SIE in the last 5 years was 0.67 million km$^2$ lower in summer and 1.17 million km$^2$ lower in winter compared to 2002–2019 (in ERA5). On an annual mean, we find that under present-day conditions, every million km$^2$ of lost sea ice may result in 11.4 Gt more SMB per year (12.9 Gt including ice shelves) (Figure A17a,b). Annually, the post-2020 trend of 219.9 Gt year$^{-1}$ over the grounded AIS (and 340.46 Gt year$^{-1}$ including ice shelves), along with 0.82 million km$^2$ less sea ice, would indicate an annual 4.3% (or 3.1% with ice shelves) increase due to sea ice loss.

During the four AR events, northerly wind speeds are significantly enhanced at 500 hPa, 850 hPa, and 10 m regardless of SIE (Figure A18). Wind speed magnitudes are similar across experiments at all heights, suggesting that sea ice loss has limited impact on dynamic AR intensification via reduced surface friction or turbulent flux changes, at least in HCLIM43 at 11 km resolution. Still, most events show increased precipitation over land and ice shelves under reduced SIC (Figure 4f–i). The most pronounced sea ice effect occurs during the March 2022 heatwave, with over 20 Gt less precipitation across four AR days when sea ice extends to ∼60°S (note that the noSIC case shows only a slight precipitation increase due to already low ice extent in the control run). In contrast, the AP heatwave in 2022 shows minimal sensitivity, likely because sea ice was already sparse (Figure A19g). Across all cases, spatial analyses reveal a complex atmospheric response to the altered SIC, caused by displaced frontal boundaries and convergence zones (Figure A19).

## Conclusions

We conclude that the recent increase in AIS mass is primarily driven by stronger westerlies and more frequent ARs. The SMB and AR increases are strongest in summer, especially on the AP and West Antarctica, associated with a positive SAM. Winter SMB has increased along most of the Antarctic coast except for West Antarctica, and we estimate that sea ice loss contributed ∼11% of the recent winter SMB increase (∼3% in summer). Together, these remote and local thermodynamic and circulation changes increased SMB enough to offset continued ice discharge losses. Our findings also show that most of the additional uptake of moisture during ARs over a sea ice-free Southern Ocean is lost locally or on ice shelves, while the strength of the higher-elevation moisture flow during ARs is largely unaffected by sea ice. Our analysis confirms the existence of a small and perhaps limited negative feedback between AIS SMB and sea ice, but more important outstanding questions on the role of climate change in enhancing precipitation from ARs in Antarctica remain, with important consequences for sea level rise estimates in the future.

# Figures

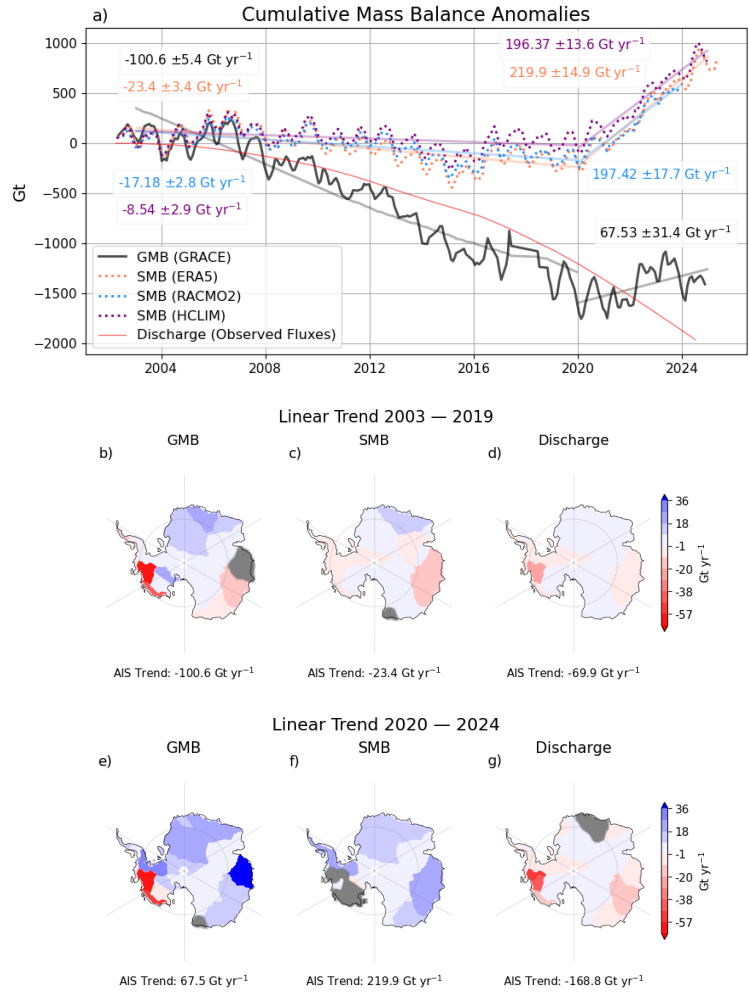

**Fig. 1**: a) Black: Monthly cumulative GMB anomalies from April 2002 to December 2024, based on GRACE (with respect to the mass as of 2011-01-01 before being subtracted from the GMB anomaly in April 2002 to show the changes since the mission launch). Dotted: Monthly cumulative SMB anomalies from April 2002 to April 2025 (ERA5), December 2024 (HCLIM43) and December 2023 (RACMO2), all with respect to the 1995-2010 mean, before subtracting their SMB from April 2002 (as for GRACE). Light straight lines indicate linear trend slopes from 2003 to 2019, and 2020 to 2024, with trends printed in respective colour (all trend slopes are significant with $p < 0.05$). Red time series: Grounding line discharge from [8] based on bed topography and velocity measurements from 1996 through to July 2024 (see Methods). Lower panel: Linear trends of GMB (b, e), SMB (c, f), and discharge (d, g) per basin for the periods 2003–2019 (upper panels) and 2020–2024 (lower panels). Basins are grayed out where the trend is not significant ($p < 0.05$). Drainage basins follow Zwally et al. (2015) [37].

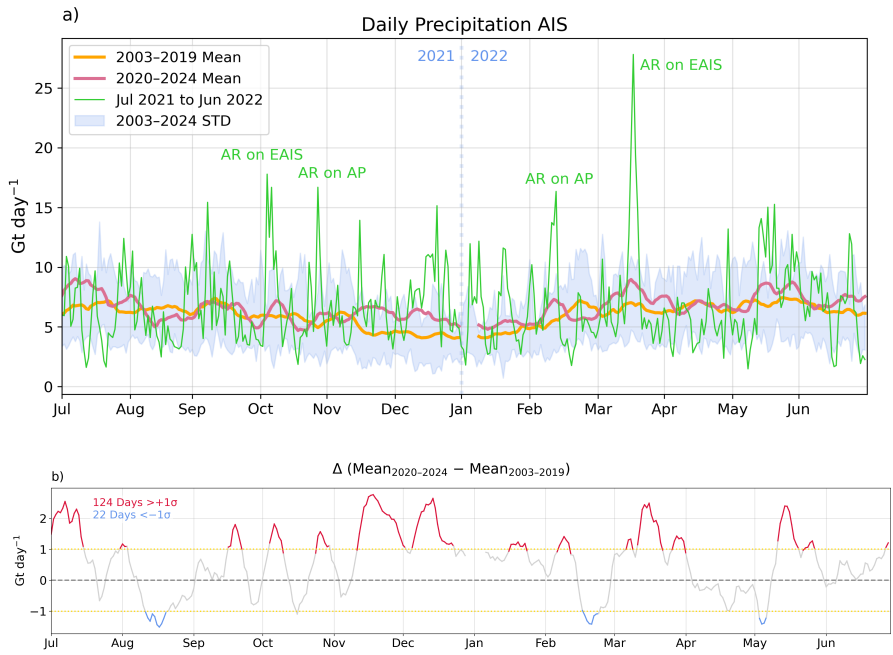

**Fig. 2**: a): Daily precipitation over the AIS throughout the year, where the green line represents the example year from Jul 2021 to Jun 2022. Blue shading indicates the daily standard deviation over the full 2003–2024 period. Red and orange line represent the mean daily precipitation (10-day rolling mean) during 2003–2019 and 2020–2024 respectively. The difference of these two lines (red minus orange) is shown in b), where red and blue coloured sections mark days where the difference is larger than the standard deviation of the difference (approximately 1 Gt day$^{-1}$), which is represented by the yellow dotted lines. 124 (22) days are significantly higher (lower) during 2020–2024 compared to 2003–2019.

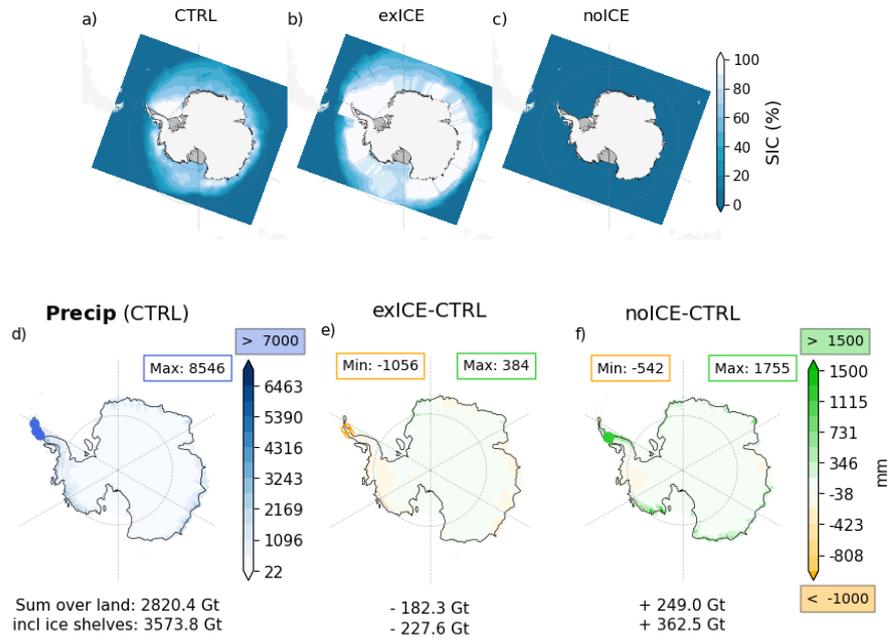

**Fig. 3**: (a-c): Mean SIC from June 2021 to July 2022 under (a) control conditions (CTRL), (b) enhanced sea ice (exICE), and (c) no sea ice (noICE) scenarios. d): Total accumulated precipitation from June 2021 to July 2022 in the control simulation, and that of the noICE (e) and exICE (f) experiments minus the control run. Numbers beneath the plots give the total amount (or difference from) for the grounded ice sheet (AIS) and including ice shelves (AIS+IS). In d), blue circles mark regions exceeding 6,000 mm with the maximum value shown in the blue rectangle. In c&d, green circles mark regions exceeding 1,500 mm and orange circles mark regions below –1,000 mm with min/max values per plot are shown in the upper rectangles; these colour scale limits were chosen to enhance the visibility of spatial patterns in the remaining areas. Respective maps for SMB are shown in Figure A11.

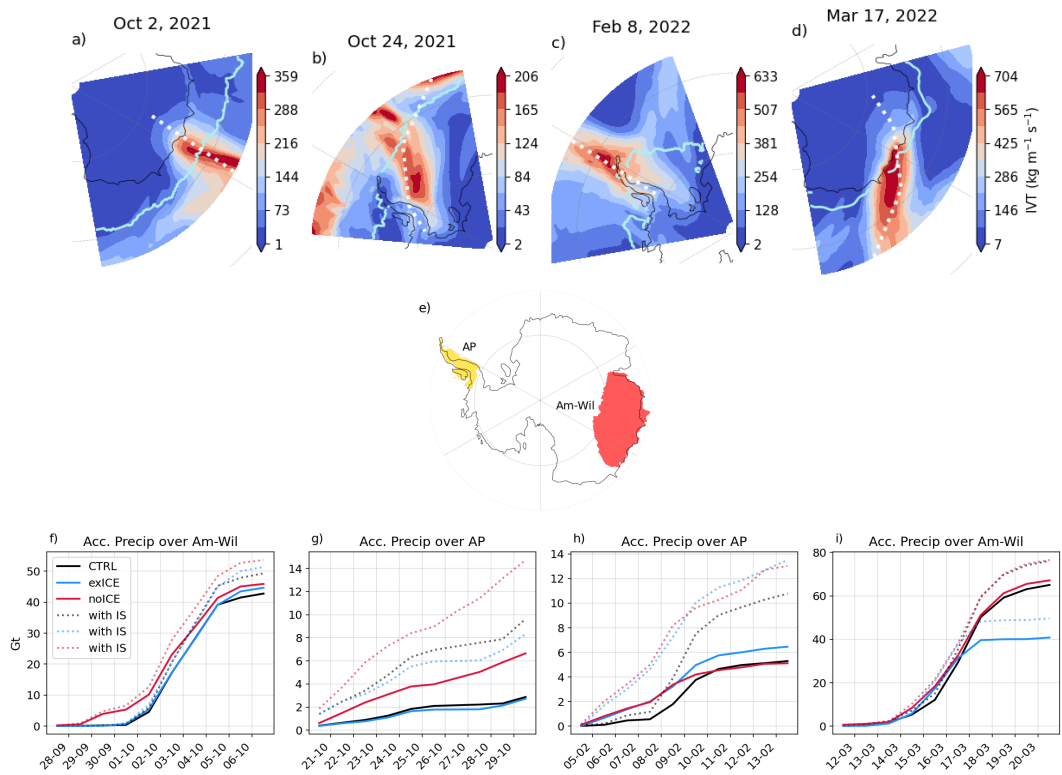

**Fig. 4**: a-d: AR events included in the 1-year sea ice experiments with HCLIM43. Here data is based on ERA5; the light blue solid line shows the sea ice edge, and the white dotted line indicates the AR axis cross-sections. e) Drainage basins which were most affected by the four ARs: the Antarctic Peninsula (AP), and Amery-and-Wilkes Land (Am-Wil). The 10-day time series below (f-i) show the accumulated precipitation for each event and experiment on the grounded ice sheet (solid lines) and that including ice shelves (dotted lines).

# Methods

## Gravitational Mass Balance

AIS mass balance can be estimated using various methods, including ice accumulation combined with ice velocity and thickness derived from satellite radar and optical imagery, altimetry (radar or laser), or gravimetry. These methods generally yield consistent results at the large scale [38, 39]. GMB trends estimates can differ greatly across studies, due to different data sources, corrections applied, slightly different time periods (e.g. Table 3 in [6] or [40]). Gravimetry data from the GRACE satellite mission, launched in 2002 (and its successor GRACE-FO, both referred to as GRACE hereafter), provides estimates of AIS mass change, after corrections for the Earth's shape [41] and glacial isostatic adjustment (GIA) [42].

We use the ESA CCI Antarctic Ice Sheet (AIS) Gravimetric Mass Balance Gridded Product (v5.0), provided by TU Dresden, derived from GRACE/GRACE-FO Level-2 monthly solutions from the Center for Space Research (CSR, RL06.3), which provides a time series of gridded as well as drainage basin-specific ice mass changes over Antarctica defined by Zwally et al. (2015) [37]. The dataset also applies a glacial isostatic adjustment (GIA) correction using the IJ05_R2 model [42] and accounts for ellipsoidal corrections [41]. It has a spatial resolution of 350 km with a 50 × 50 km grid and covers the period April 2002 to December 2024. Ice mass changes are referenced to 2011-01-01 based on a linear, periodic (annual and semi-annual), and quadratic fit to monthly solutions over 2002-08 to 2016-08. We subtracted all GMB anomalies from the GMB anomaly of April 2002 to show the mass loss since 2002 in line with the SMB data (Figure 1a).

## Surface Mass Balance

We use ERA5 meteorological precipitation and evaporation fields to estimate SMB from 1985 to April 2025, calculated as precipitation minus evaporation. Although additional processes influence the final SMB, this provides a reasonable approximation when compared with more advanced SMB estimates [43]. We also use downscaled SMB (precipitation-evaporation) simulations based on HCLIM43 [36] in its AROME (11 km) configuration.

To compare ERA5 SMB to more sophisticated SMB calculations, we further use updated RACMO2.4p (RACMO2 in manuscript) SMB data on 11 km grids from 1985 to 2023 [44]. RACMO2 estimates SMB as the sum of snowfall, rainfall, sublimation, drifting snow erosion, and meltwater runoff, using a multi-layer snow scheme that includes snow densification, melt percolation, and refreezing. The model is driven by ERA5 reanalysis at the lateral boundaries and incorporates updated IFS physics, including prognostic precipitation types and a spectral snow albedo scheme coupled to the TARTES radiative transfer model. Snow processes specific to polar conditions, such as blowing snow and superimposed ice formation, are explicitly represented.

SMB anomalies for all SMB data are derived by subtracting the 1995-2010 mean (as 1995 was the first time step for HCIM).

**Atmospheric Rivers**

Two AR algorithms were used to evaluate changes in AR frequency and AR precipitation: ANTIS-AR and EDARA [45], which both use IVT based on ERA5 data. We use the vertical integrals of eastward ($F_u$) and northward ($F_v$) water vapour fluxes, which represent the total column-integrated moisture flux (in $\text{kg}\,\text{m}^{-1}\,\text{s}^{-1}$) from the surface to the top of the atmosphere (calculated on model levels that follow terrain). IVT is then calculated as

$$\text{IVT} = \sqrt{F_u^2 + F_v^2} \tag{1}$$

$F_u$ and $F_v$ are derived internally by ECMWF from model-level winds and specific humidity as

$$F_u, F_v = -\frac{1}{g} \int_{p_{\text{sfc}}}^{p_{\text{top}}} (u, v)\, q\, \mathrm{d}p \tag{2}$$

where $u$ and $v$ are the horizontal wind components, $q$ is specific humidity, $p$ is pressure, and $g$ is gravitational acceleration.

The ANTIS-AR algorithm was specifically developed for this study to identify extreme zonal and meridional ARs that make landfall on the ice sheet and/or shelves. It detects contiguous regions where IVT exceeds the $95_{\text{th}}$ percentile of the monthly climatological $\text{IVT}_{\text{u,v}}$ at each grid point, with a minimum IVT threshold of 40 kg $\text{m}^{-1}\,\text{s}^{-1}$. Identified regions must have a length-to-width ratio of at least 2:1 [46] and a minimum length of 1,300 km. We reduced this threshold from the more common global definitions of 1,500 or 2,000 km, as we only evaluated IVT fields south of 60°S (i.e. ARs extending farther north are truncated). ARs were detected on the original 25km grid and daily time steps. A 1-year sensitivity analysis for 2024 revealed a negligible increase in AR frequency (from 6.24 to 6.36 AR days $\text{yr}^{-1}$ on average) if detected on 12-hour time steps.

EDARA [45] is a global ERA5-based AR dataset that follows the detection framework of Guan et al. (2015) [47]. It is not specifically optimized for polar regions (e.g. it uses a relatively high IVT threshold of 100 kg $\text{m}^{-1}\,\text{s}^{-1}$), but provides a valuable baseline to compare AR changes detected using our ANTIS-AR algorithm. EDARA detects ARs that exceed the $85_{\text{th}}$ percentile of $\text{IVT}_{\text{u,v}}$, and is therefore technically less strict than ANTIS-AR. However, because the absolute IVT threshold of 100 kg $\text{m}^{-1}\,\text{s}^{-1}$ must also be exceeded (which is rarely the case over Antarctica even during strong ARs), EDARA tends to underestimate AR frequency and extent over the AIS, but detects more ARs over the surrounding Southern Ocean. The used EDARA-ARs in this study are based on the mtARget-v3 algorithm, which improves the identification of zonally oriented ARs. Since this study focuses on ARs affecting land and ice shelves, only EDARA ARs that reach at least one grid cell of land or ice shelf are considered.

## HCLIM43 experiments

The Pan-Antarctic sea ice experiment simulations were performed over the Antarctic region using the non-hydrostatic regional climate model HCLIM-AROME, version 43 [36], at 11 km resolution. HCLIM43-AROME is a convection-permitting, nonhydrostatic configuration designed to explicitly resolve deep convection processes. It uses a one-way nesting setup and was run with 65 vertical levels. The model includes thermodynamic sea ice scheme (SICE), a multilayer snow scheme (ISBA-3L) and advanced parameterizations for turbulence, microphysics, and shallow convection. An overview of the model physics used in HCLIM43-AROME is summarized below, and the domain boundaries are shown in Figure 3a.

- **Dynamics:** Nonhydrostatic [48]
- **Radiation:** RRTM for longwave, SW6 for shortwave
- **Turbulence:** HARATU scheme [49, 50]
- **Microphysics:** ICE3-OCND2 [51, 52]
- **Shallow convection:** EDMFm [50, 53]
- **Deep convection:** Explicitly resolved (no parameterization)
- **Cloud scheme:** [54]
- **Orographic drag:** Not included
- **Sea ice:** SICE module [55]
- **Snow:** Multilayer ISBA-3L scheme [56]

HCLIM43-AROME was driven using ERA5 reanalysis data, including temperature, zonal and meridional winds, specific humidity, sea ice concentration, sea surface temperature, and surface pressure, updated every three hours. Nudging was applied to air temperature, divergence and vorticity above 850 hPa, with a length scale of approximately 800 km. Moisture fields, such as water vapour, were excluded. The downscaling experiment was run from July 2021 to June 2022, with one month of spin-up (Torres-Alavez et al.; in preparation).

## SAM and ENSO

To analyse the impact of variability of planetary climate modes we used the SAM and ENSO indices provided at http://www.nerc-bas.ac.uk/icd/gjma/sam.html and https://ds.data.jma.go.jp/tcc/tcc/products/elnino/index/respectively. As there are different types of El Niños, we focused on the two main ones: El Niño3 and El Niño4, which we averaged in this study as the time series are very comparable. We also did sensitivity test of SMB and El Niño3 and El Niño4 indices separately which resulted in very similar results (e.g. Figure A3).

**Supplementary information.**

- Funding
  R. Mottram's participation in this research was supported by Ocean Cryosphere Exchanges in Antarctica: Impacts on Climate and the Earth system, OCEAN ICE, which is funded by the European Union, Horizon Europe Funding Programme for research and innovation under grant agreement Nr. 101060452, 10.3030/101060452. OCEAN ICE contribution number 34. J.A. Torres-Alvarez was supported by the ESA project Polar Ice Sheets in Climate models and earth Observation (PISCO) ESA Contract No. 4000148150/25/I-LR. We acknowledge the generous help and support of the ESA climate change initiative for the Antarctic Ice Sheet and for Sea Ice, without whose climate data records this study would not have been possible. We also acknowledge the support of the European Centre for Medium Range Weather Forecasting for computational resources in special project spdkmott. Ideas in this manuscript were initially provoked during M. Kolbe's PhD visit to DMI and subsequently developed further during a research stay at the University of Canterbury, funded under the project number ALWPP.2019.003 of the research programme NPP which is (partly) financed by the Dutch Research Council (NWO).

- Data availability
  Data for the three different HCLIM43 experiments have been uploaded to https://ensemblesrt3.dmi.dk/data/prudence/temp/JAT/Marlen/year/ (CTL denotes the control simulation, WICE represents the idealised experiment with added sea ice, and NICE the experiment with completely removed sea ice; subfolders are ordered by month from June 2021 to July 2022). Gravitational mass balance data was sourced from https://data1.geo.tu-dresden.de/ais_gmb/, where GRACE/GRACE-FO L2 monthly solutions are provided by the Center for Space Research (CSR RL06.2). RACMO2-based SMB is provided at https://zenodo.org/records/14217232. ERA5 meteorological fields were retrieved via the Climate Data Store (CDS) infrastructure https://cds.climate.copernicus.eu/datasets/reanalysis-era5-complete?tab=overview. Grounding line discharge estimates are available at https://doi.org/10.5281/zenodo.10051893 [8]. Indices for SAM and ENSO (El Niño 3&4 SST anomalies) are provided at http://www.nerc-bas.ac.uk/icd/gjma/sam.html and https://ds.data.jma.go.jp/tcc/tcc/products/elnino/index/ respectively.

- Code availability
  Python scripts for the AR detection algorithm for ANTIS-AR as well as the code for filtering ARs from EDARA that reach the Antarctic ice sheet or shelves can be accessed via https://doi.org/10.5281/zenodo.15645442. The EDARA IVT and AR detection scripts by Mo (2024) are available at https://www.frdr-dfdr.ca/repo/dataset/b1798e59-b38e-4a83-ab88-12d0a8aca28f.

- Competing Interest Information
  The authors have no competing interests to declare that are relevant to the content of this article.

- Author contribution
  All authors contributed to the development of the ideas and methods presented in this study. J.A. Torres-Alvarez conducted the HCLIM43 simulations, and M. Kolbe performed the atmospheric river detection. M. Kolbe carried out the data analysis and wrote the main manuscript, with input from R. Mottram, J.A. Torres-Alvarez, R. Bintanja, E. van der Linden, and M. Katurji.

# Appendix A  Supplementary Figures

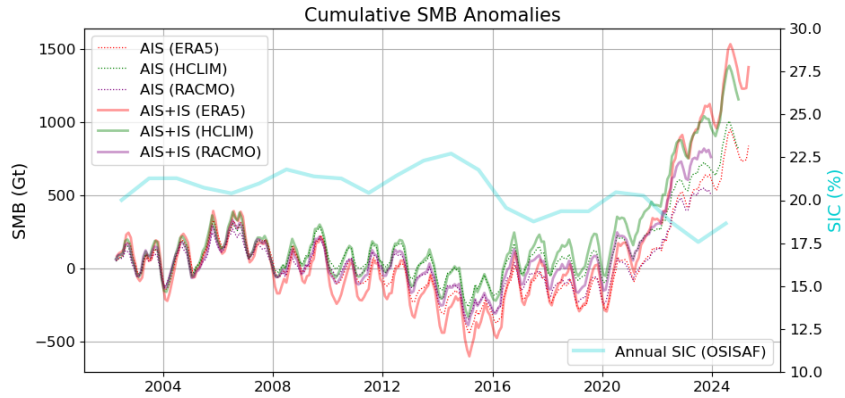

**Fig. A1**: SMB as in Figure 1a but differentiating between SMB over the grounded ice sheet (AIS; dotted lines) and the same including ice shelves (AIS+IS; solid lines), with annual SIC based on OSISAF shown in blue.

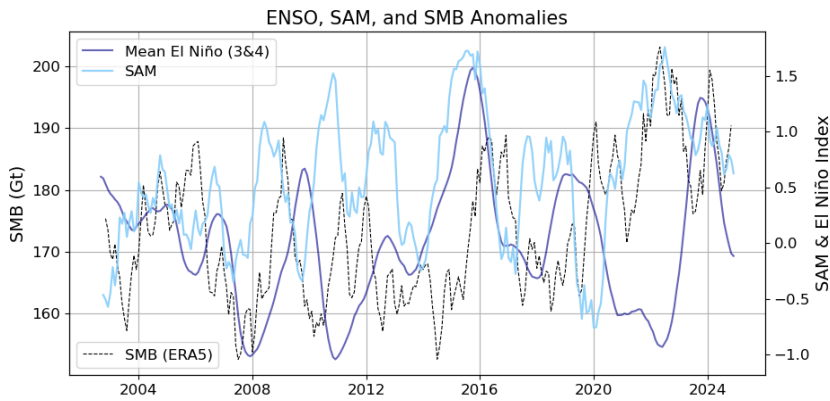

**Fig. A2**: Black dashed lined: Monthly SMB anomalies based on ERA5 (with respect to the 1995-2010 mean) over the AIS during the GRACE period (2003–2024). Light blue line and dark blue line represent the time series for the SAM index and the mean of the El Niño3 and El Niño4 indices respectively (the two individual El Niño time series were very similar).

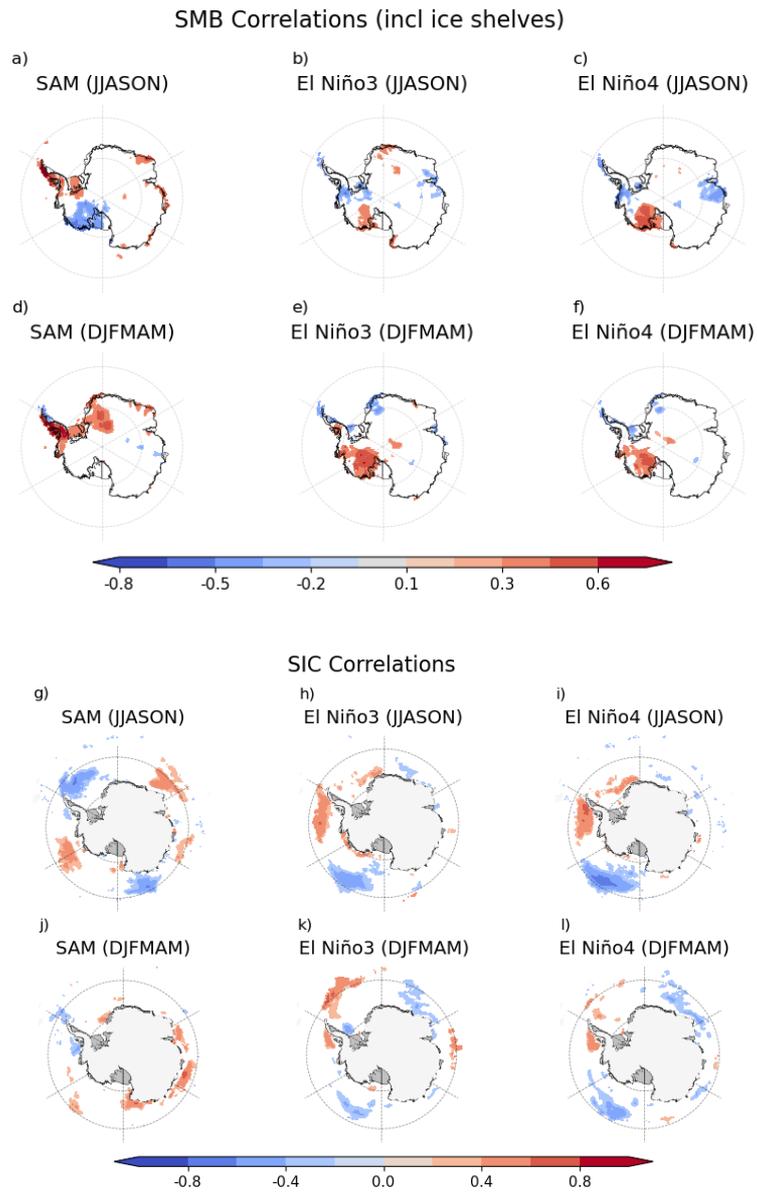

**Fig. A3**: a–f: Correlations of the SAM (a), El Niño3 (b) and El Niño4 (c) indices with gridded SMB including ice shelves during winter (JJASON). d–f: Same but for summer (DJFMAM). Grid points with low significance correlations (p-values ¿ 0.05) are masked out (white). g–l: Correlations of the SAM (g), El Niño3 (h) and El Niño4 (i) indices with gridded SIC during winter (JJASON). j–l: Same but for summer (DJFMAM).

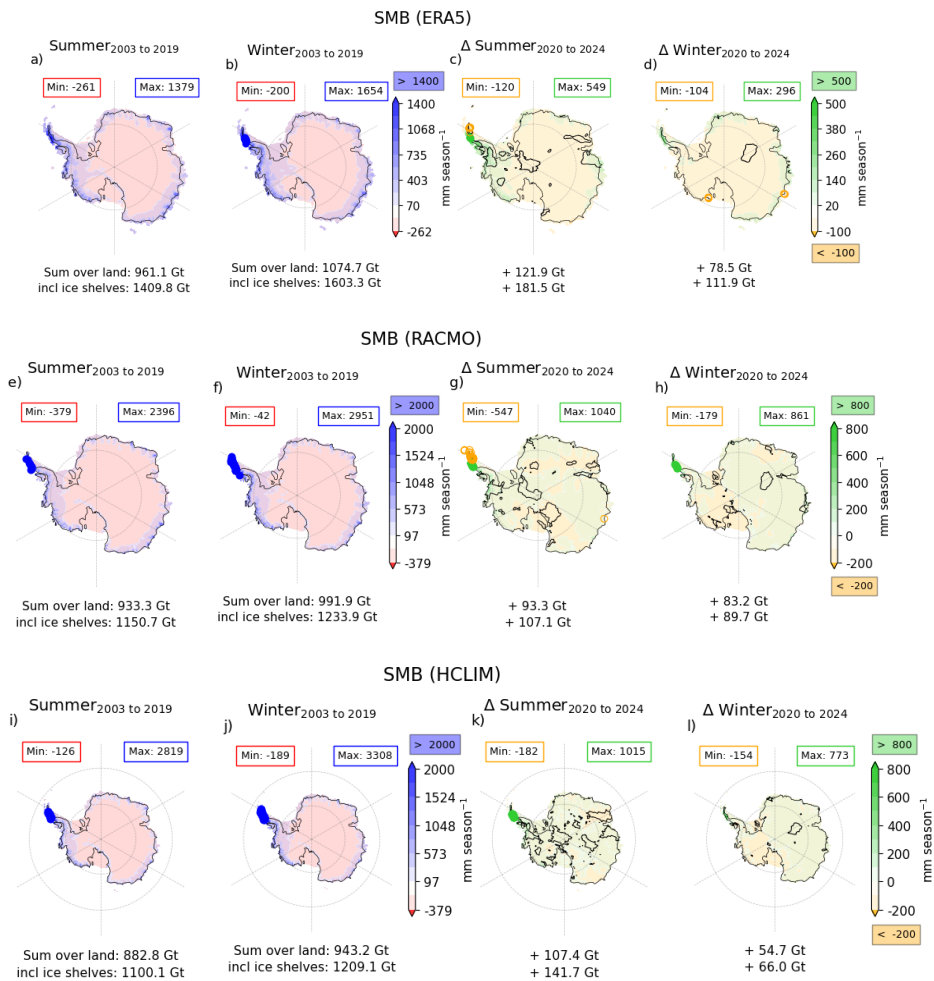

**Fig. A4**: Mean sum of SMB in ERA5 during 2003–2019 in summer (DJFMAM; a) and winter (JJASON; b). c) and d) illustrate the difference of the five recent years (2020–2024 mean) compared to the 2003–2019 mean, i.e. blue (red) areas mark regions with higher (lower) SMB. Black contours mark areas where the difference was significant based on Welch's t-test, which accounts for differences in population size. In a&b, blue circles mark regions exceeding 1,400 mm per season. In c&d, green circles mark regions exceeding 500 mm per season and orange circles mark regions below –100 mm per season. e-h: Same as in a-d but for RACMO2. e-h: Same as in a-d but for HCLIM43 (precipitation minus evaporation). Note the extended colour scales compared to ERA5 due to higher resolution (and in RACMO also a more accurate SMB).

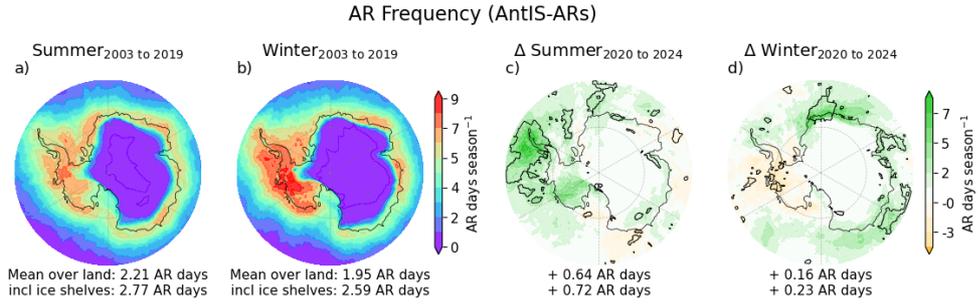

**Fig. A5**: Mean AR frequency based on the ANTIS-AR algorithm (see Methods) during 2003–2019 in summer (DJFMAM; a) and winter (JJASON; b). c) and d) illustrate the difference of the five recent years (2020–2024 mean) compared to the 2003–2019 mean, i.e. green (orange) areas mark regions with higher (lower) AR frequency. Black contours mark areas where the difference was significant based on Welch's t-test, which accounts for differences in population size.

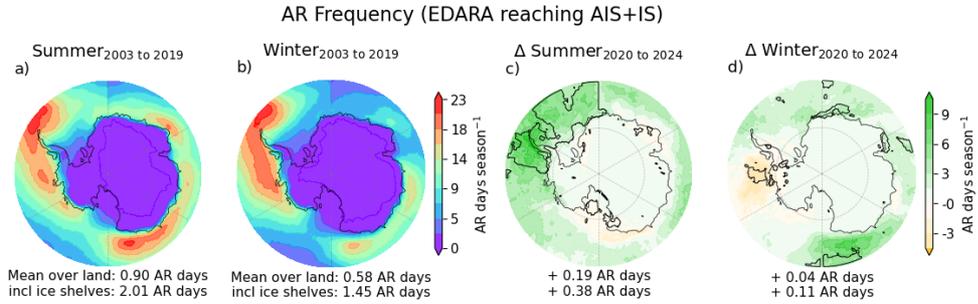

**Fig. A6**: As in Figure A5, but for AR frequency based on the adapted EDARA algorithm, where only ARs that reach the ice sheet or ice shelves are retained (see Methods). Because the detection algorithm is not adjusted for polar regions and uses a minimum IVT threshold of $100 \, \text{kg m}^{-1} \, \text{s}^{-1}$, fewer ARs are identified over the AIS compared to the ANTIS-AR algorithm (and more over the ocean due to the lower $85_{\text{th}}$ percentile).

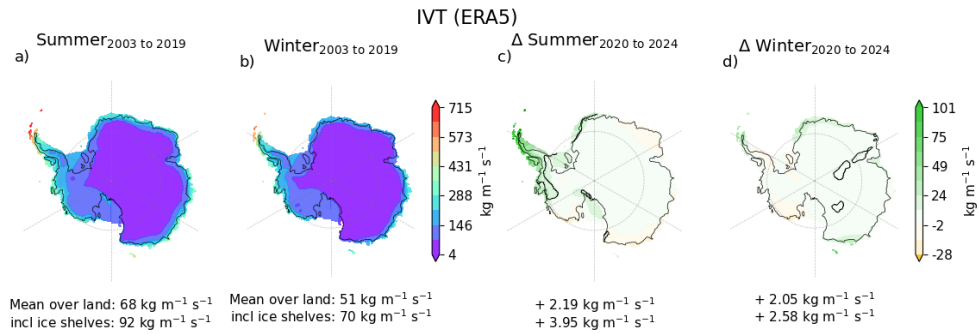

**Fig. A7**: Mean of IVT over the AIS and ice shelves during 2003–2019 in summer (DJFMAM; a) and winter (JJASON; b) over the AIS and ice shelves. c) and d) illustrate the difference of the five recent years (2020–2024 mean) compared to the 2003–2019 mean, i.e. red (blue) areas mark regions with higher (lower) AR frequency. Black contours mark areas where the difference was significant based on Welch's t-test, which accounts for differences in population size.

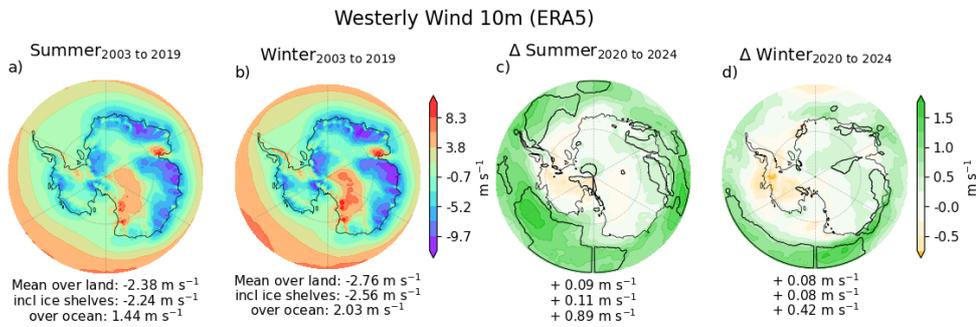

**Fig. A8**: As in Figure A7, but for the mean zonal wind at 10m including oceanic areas, where positive values mark westerly winds. Zonal winds at 850 hPa and 500 hPa show the same pattern with similar regions of significance (plots can be provided upon request).

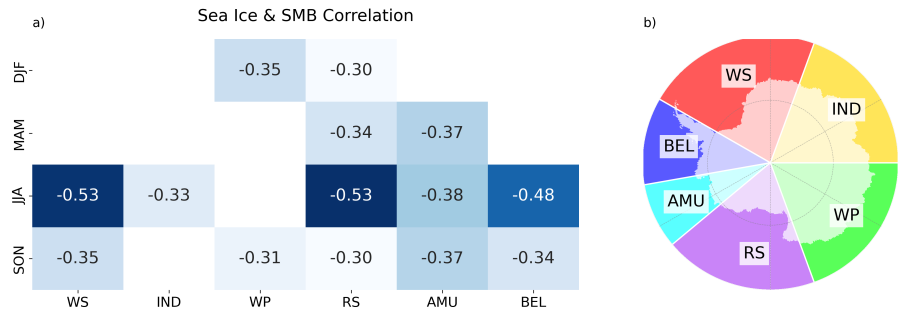

**Fig. A9**: a) Correlations of basin-wide SMB and adjacent ocean basins for SIC. The basins are shown in b, where the transparent land/ice shelf areas represent the region where SMB anomalies are summed, and correlated to the respective adjacent ocean basins. WS: Weddell Sea, IND: Indian Ocean, WP: West Pacific, RS: Ross Sea, AMU: Amundsen Sea, BEL: Bellingshausen Sea. The correlations per grid point are shown in Figure A10.

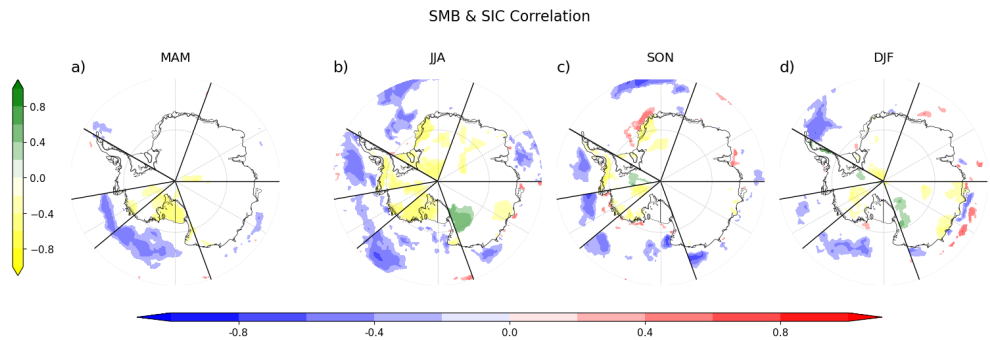

**Fig. A10**: As in Figure A9a, but showing correlations per sector of total SMB with gridded SIC (blue to red), and of mean SIC with gridded SMB (including ice shelves; yellow to green). Yellow patterns over land and blue over ocean thus indicate co-occurring high SMB and low SIC. Grid points with low significance correlations (p-values 0.05) are masked out (white).

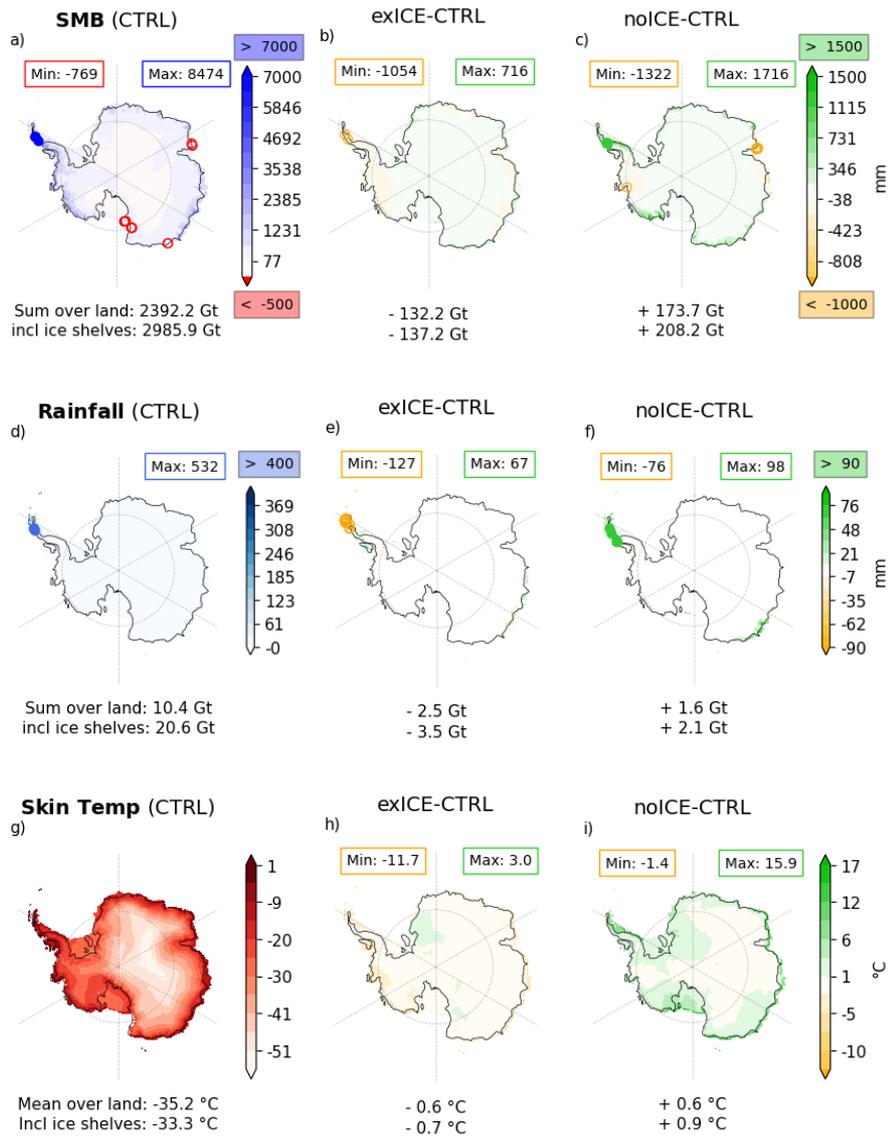

**Fig. A11**: As in Figure 3d–f, but for SMB (a–c), rainfall (d-f), and skin temperature (g-i).

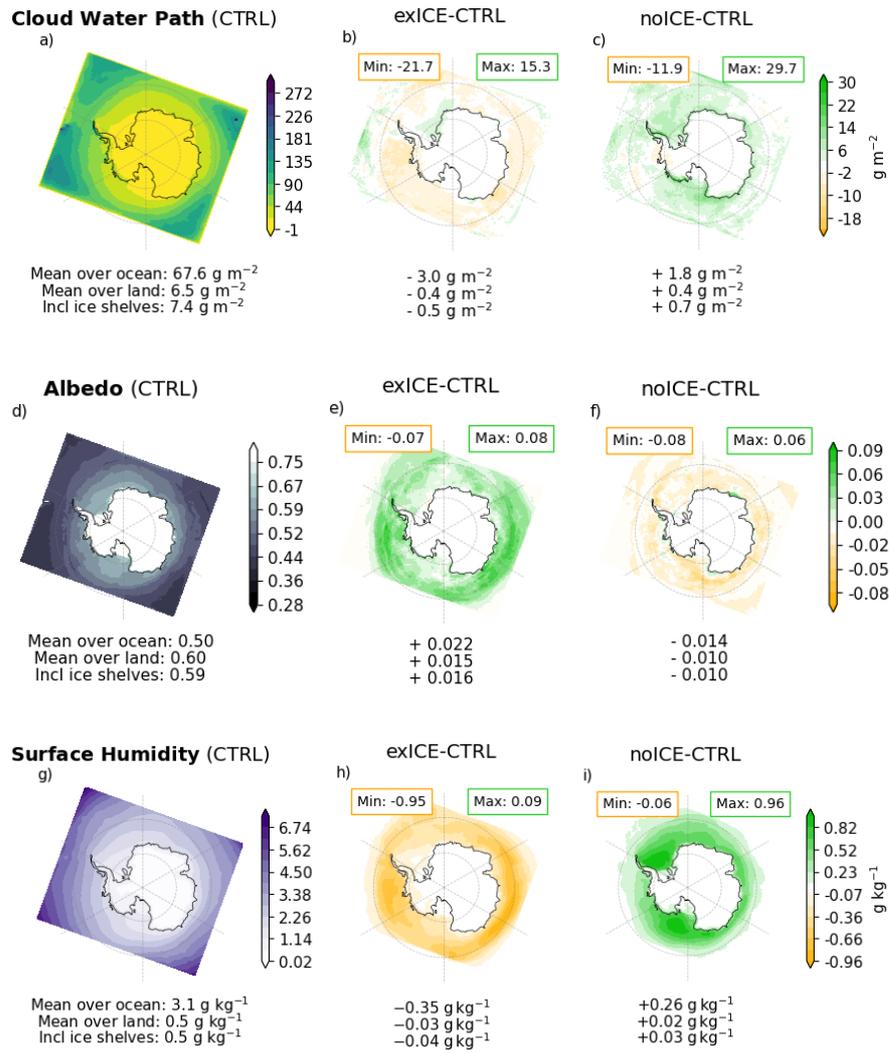

**Fig. A12**: As in Figure 3d–f, including the Southern Ocean, but for cloud water path (a-c), surface albedo (d-f) and surface humidity (g-i).

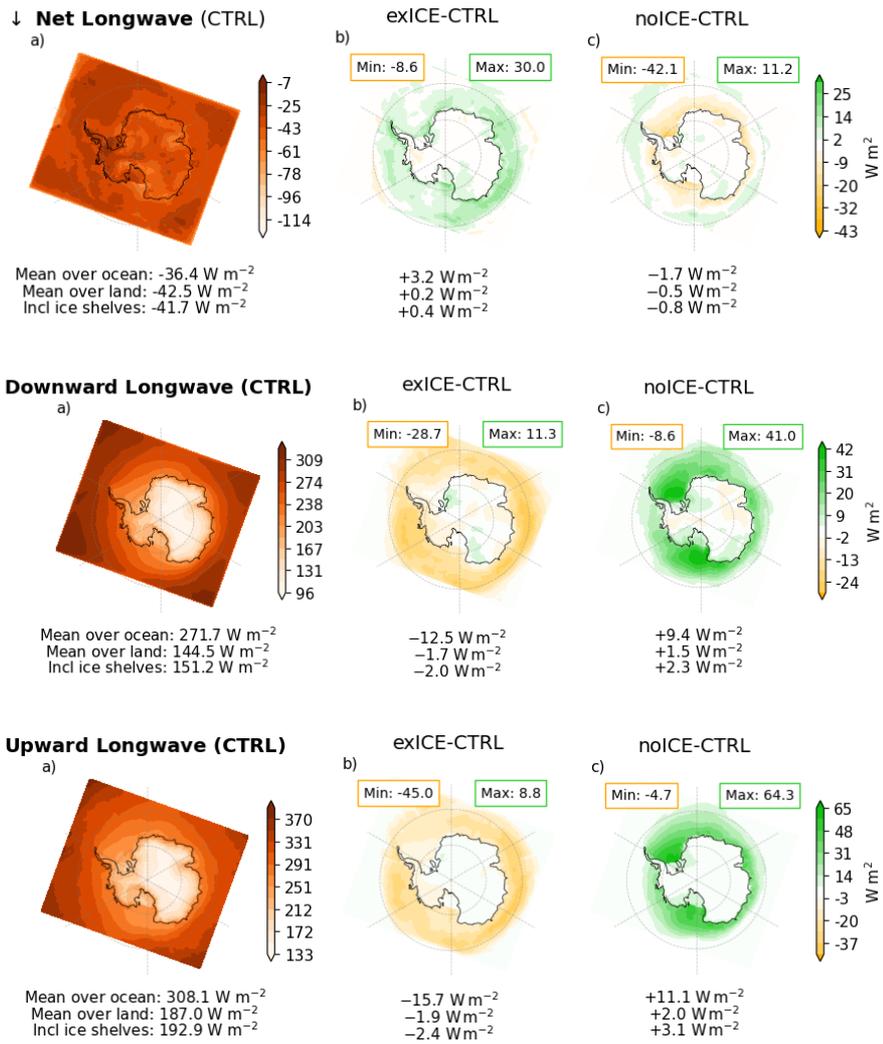

**Fig. A13**: As in Figure 3d–f, including the Southern Ocean, but for net (a-c), downward (d-f), and upward (g-i) longwave radiation (all downward positive).

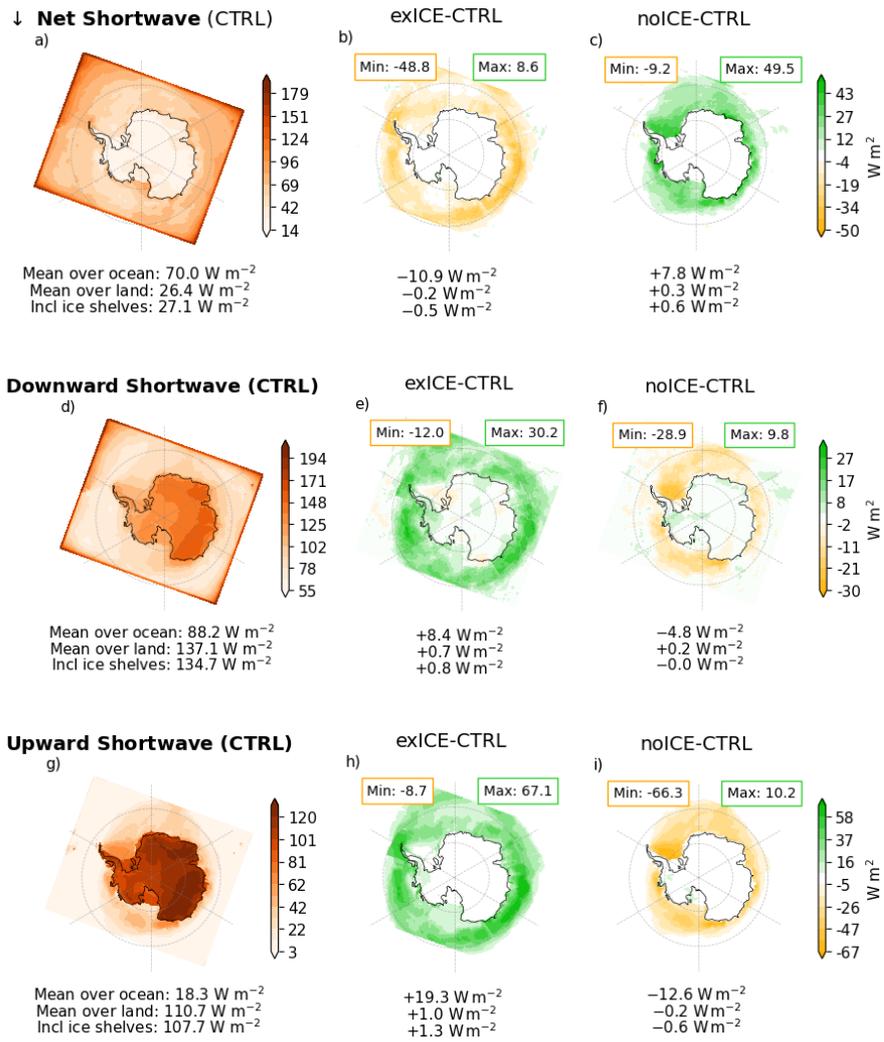

**Fig. A14**: As in Figure 3d–f, including the Southern Ocean, but for net (a-c), downward (d-f), and upward (g-i) shortwave radiation (all downward positive).

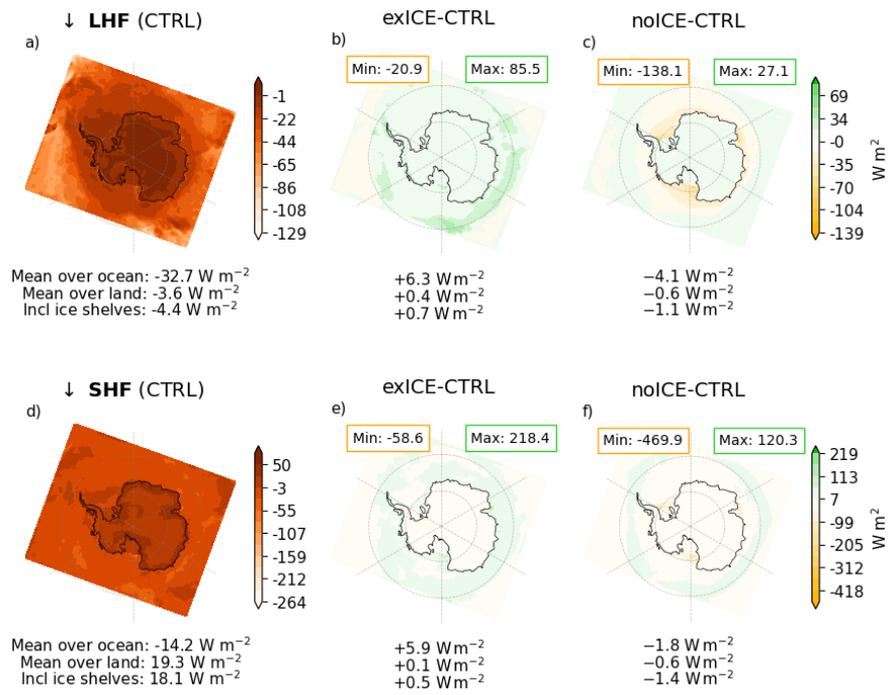

**Fig. A15**: As in Figure 3d–f, including the Southern Ocean, but for latent (a-c) and sensible (d-f) turbulent heat fluxes (both downward positive).

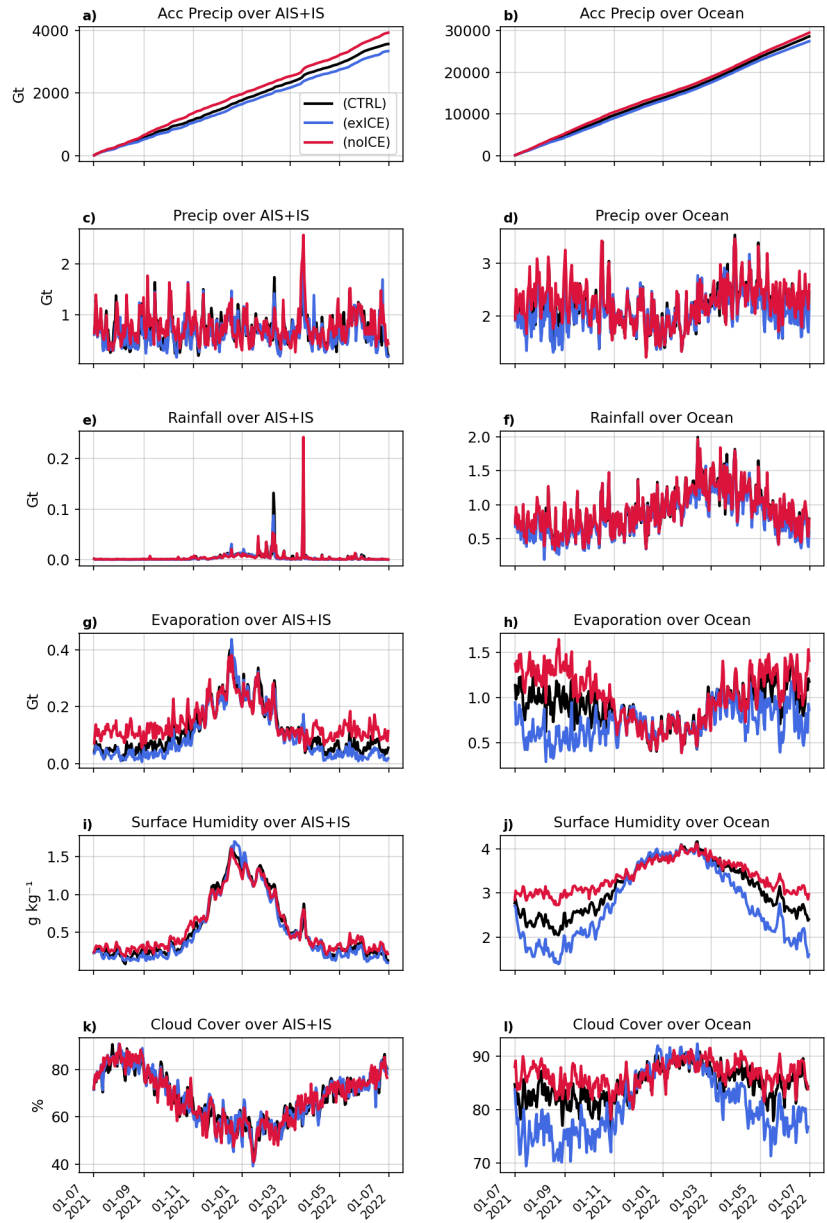

**Fig. A16**: Weighted mean or sum of selected variables averaged over the AIS+IS (left) and Southern Ocean (right) from July 2021 to June 2022 in the three experiments (CTRL, exICE, noICE). Results over the AIS excluding ice shelves are very similar to the left column. Continued on following page.

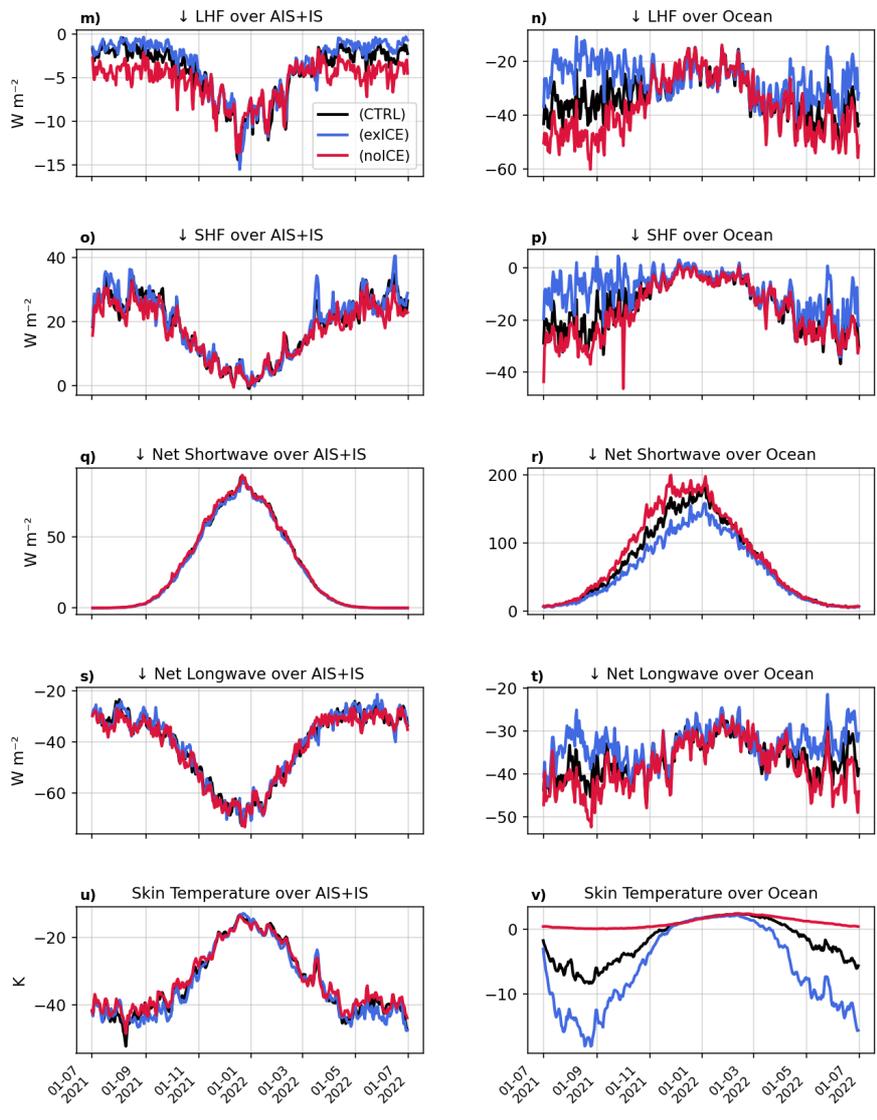

**Fig. A16**: (continued, see previous page)

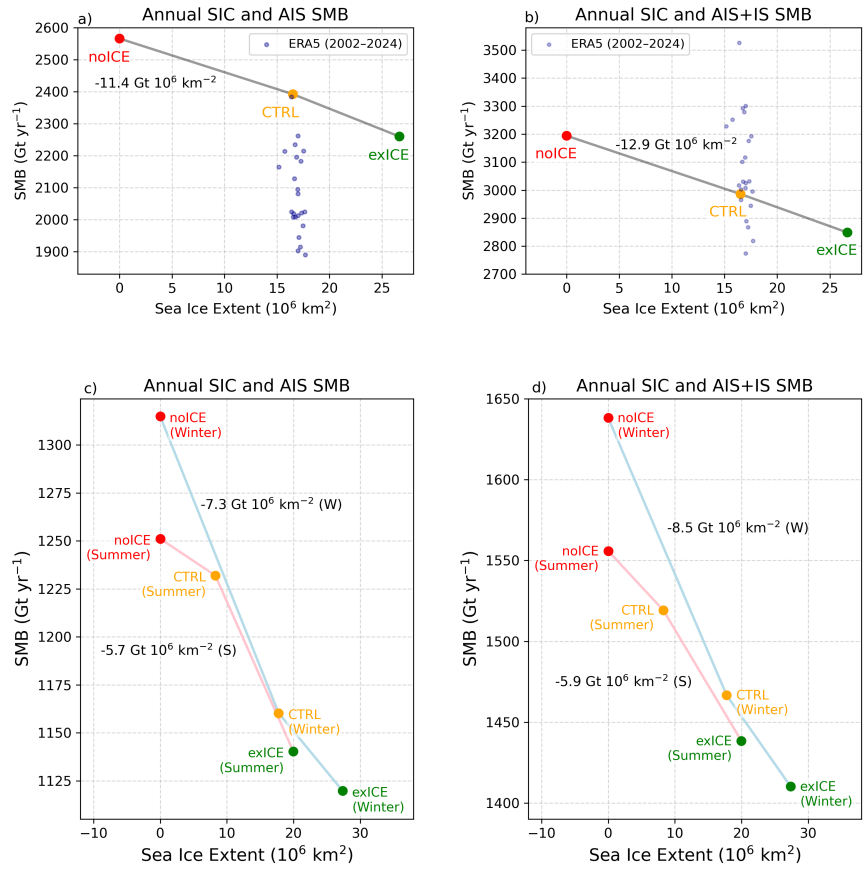

**Fig. A17**: SIE and SMB relationships for the 1-year noICE, CTRL, and exICE experiments (07/2021 to 06/2022) with HCLIM43. Top: annual mean and accumulated SMB for the grounded AIS (a) and including ice shelves (b). Bottom: Differences between summer and winter for the AIS (c) and including ice shelves (d). In a) and b), all annual years from 2002-2024 based on ERA5 are added (which is used to drive the HCLIM43 experiments). SMB is approximated as precipitation minus evaporation.

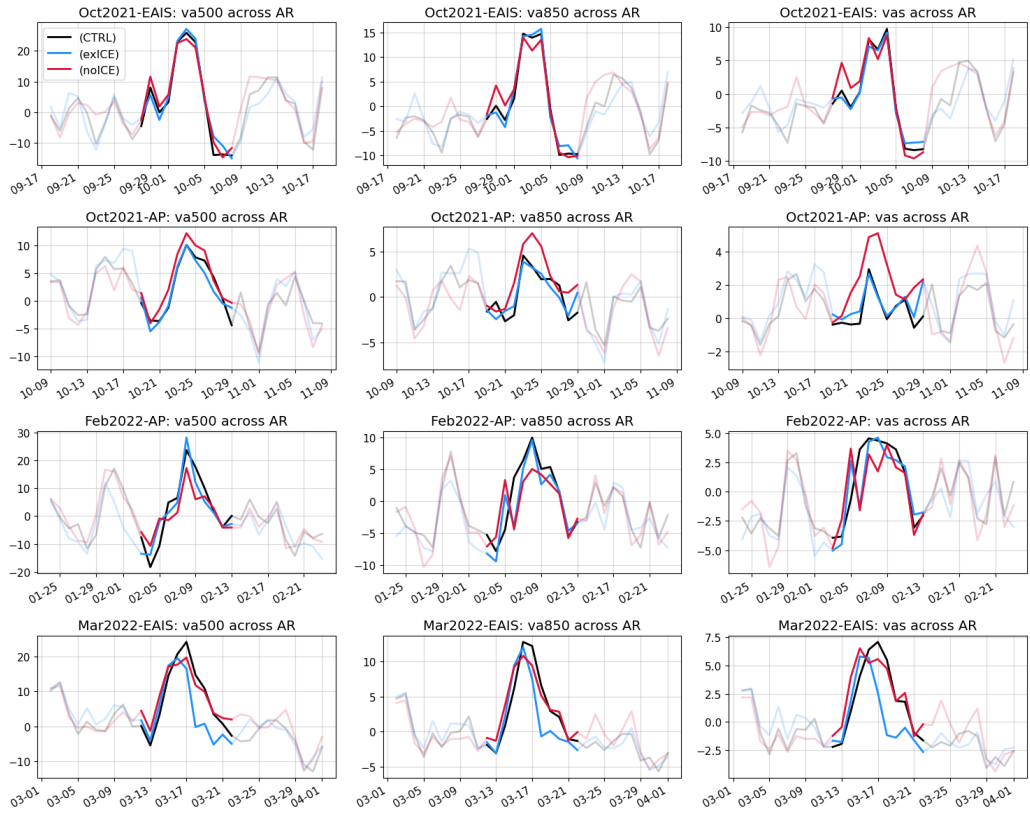

**Fig. A18**: Northerly winds of the 3 experiment simulations at 500 hPa (left), 500 hPa (middle), and 10m (right) averaged over the AR cross section of the October AR on the EAIS (a-c), the October AR on the AP (d-f), the February AR on the AP (g-i), and the March AR on the EAIS (j-l).

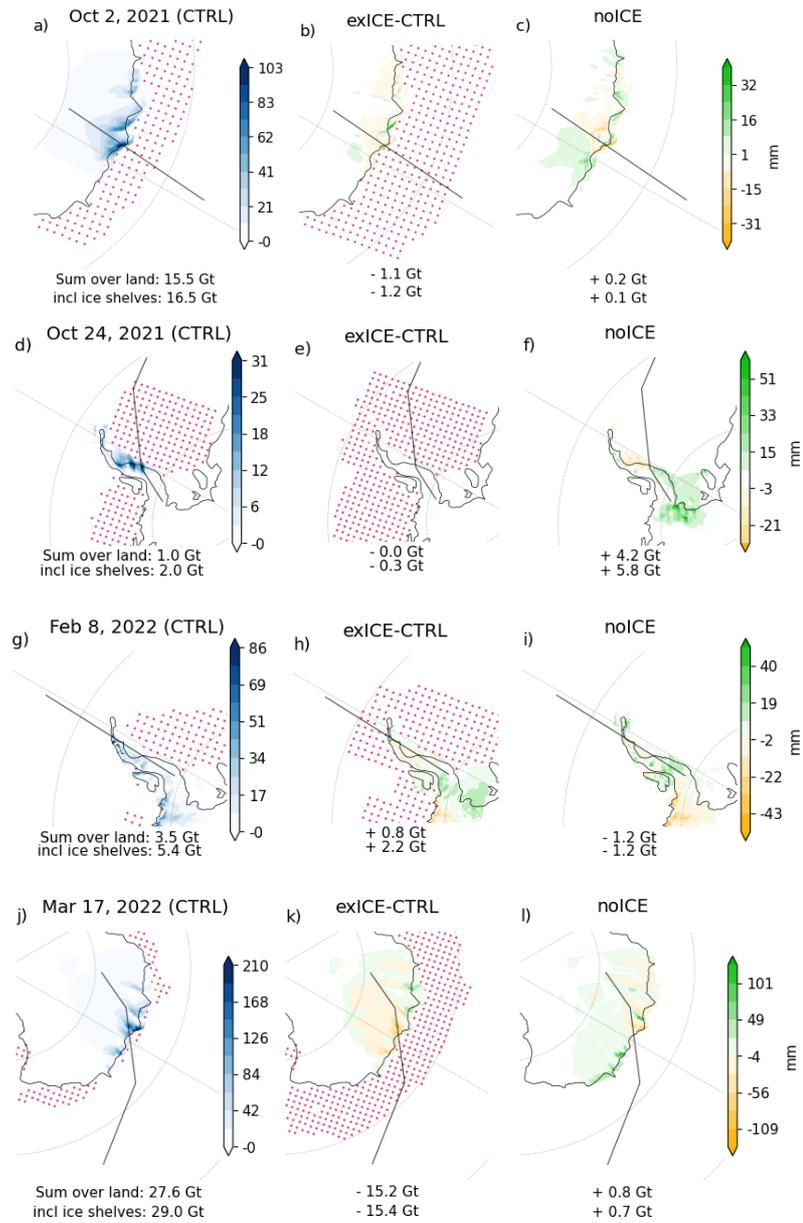

**Fig. A19**: Precipitation on the main AR day of the four events (a,d,g,j) and the difference of the exICE and noICE experiments to the CTRL. Black lines indicate the AR axes (as in Figure 4a-d), and SIC exceeding 15% are marked with red dots.

# References

[1] Purich, A. & England, M. H. Projected impacts of antarctic meltwater anomalies over the twenty-first century. *Journal of Climate* **36**, 2703–2719 (2023).

[2] Li, Q., England, M. H., Hogg, A. M., Rintoul, S. R. & Morrison, A. K. Abyssal ocean overturning slowdown and warming driven by antarctic meltwater. *Nature* **615**, 841–847 (2023).

[3] Ferrari, R. *et al.* Antarctic sea ice control on ocean circulation in present and glacial climates. *Proceedings of the National Academy of Sciences* **111**, 8753–8758 (2014).

[4] Berk, J. v. d., Drijfhout, S. & Hazeleger, W. Circulation adjustment in the arctic and atlantic in response to greenland and antarctic mass loss. *Climate Dynamics* **57**, 1689–1707 (2021).

[5] Wang, W., Shen, Y., Chen, Q. & Wang, F. Unprecedented mass gain over the antarctic ice sheet between 2021 and 2022 caused by large precipitation anomalies. *Environmental Research Letters* **18**, 124012 (2023).

[6] Zhang, R., Xu, M., Che, T., Guo, W. & Li, X. Ice sheet mass changes over antarctica based on grace data. *Remote Sensing* **16**, 3776 (2024).

[7] Ekaykin, A. A., Veres, A. N. & Wang, Y. Recent increase in the surface mass balance in central east antarctica is unprecedented for the last 2000 years. *Communications Earth & Environment* **5**, 200 (2024).

[8] Davison, B. J., Hogg, A. E., Slater, T. & Rigby, R. Antarctic ice sheet grounding line discharge from 1996 through 2023. *Earth System Science Data Discussions* **2023**, 1–35 (2023).

[9] Li, J. *et al.* Unraveling the contributions of atmospheric rivers on antarctica crustal deformation and its spatiotemporal distribution during the past decade. *Geophysical Journal International* **235**, 1325–1338 (2023).

[10] Dalaiden, Q., Goosse, H., Lenaerts, J. T., Cavitte, M. G. & Henderson, N. Future antarctic snow accumulation trend is dominated by atmospheric synoptic-scale events. *Communications Earth & Environment* **1**, 62 (2020).

[11] Nash, D., Waliser, D., Guan, B., Ye, H. & Ralph, F. M. The role of atmospheric rivers in extratropical and polar hydroclimate. *Journal of Geophysical Research: Atmospheres* **123**, 6804–6821 (2018).

[12] Wille, J. D. *et al.* Atmospheric rivers in antarctica. *Nature Reviews Earth & Environment* 1–15 (2025).

[13] Gorodetskaya, I. V. *et al.* Record-high antarctic peninsula temperatures and surface melt in february 2022: a compound event with an intense atmospheric river. *npj climate and atmospheric science* **6**, 202 (2023).

[14] Zou, X. *et al.* Strong warming over the antarctic peninsula during combined atmospheric river and foehn events: contribution of shortwave radiation and turbulence. *Journal of Geophysical Research: Atmospheres* **128**, e2022JD038138 (2023).

[15] Wille, J. D. *et al.* The extraordinary march 2022 east antarctica "heat" wave. part i: observations and meteorological drivers. *Journal of Climate* **37**, 757–778 (2024).

[16] Wille, J. D. *et al.* The extraordinary march 2022 east antarctica "heat" wave. part ii: impacts on the antarctic ice sheet. *Journal of Climate* **37**, 779–799 (2024).

[17] Blanchard-Wrigglesworth, E., Cox, T., Espinosa, Z. I. & Donohoe, A. The largest ever recorded heatwave—characteristics and attribution of the antarctic heatwave of march 2022. *Geophysical Research Letters* **50**, e2023GL104910 (2023).

[18] Kolbe, M. *et al.* Model performance and surface impacts of atmospheric river events in antarctica. *Discover Atmosphere* **3**, 4 (2025).

[19] Groh, A. & Horwath, M. Antarctic ice mass change products from grace/grace-fo using tailored sensitivity kernels. *Remote Sensing* **13**, 1736 (2021).

[20] Velicogna, I. *et al.* Continuity of ice sheet mass loss in greenland and antarctica from the grace and grace follow-on missions. *Geophysical Research Letters* **47**, e2020GL087291 (2020).

[21] Loomis, B., Luthcke, S. & Sabaka, T. Regularization and error characterization of grace mascons. *Journal of geodesy* **93**, 1381–1398 (2019).

[22] Mottram, R. *et al.* What is the surface mass balance of antarctica? an intercomparison of regional climate model estimates. *The Cryosphere* **15**, 3751–3784 (2021).

[23] Nicola, L., Notz, D. & Winkelmann, R. Revisiting temperature sensitivity: how does antarctic precipitation change with temperature? *The Cryosphere* **17**, 2563–2583 (2023).

[24] Medley, B. *et al.* Temperature and snowfall in western queen maud land increasing faster than climate model projections. *Geophysical Research Letters* **45**, 1472–1480 (2018).

[25] Wang, J. *et al.* The impacts of combined sam and enso on seasonal antarctic sea ice changes. *Journal of Climate* **36**, 3553–3569 (2023).

[26] Porter, S. E., Parkinson, C. L. & Mosley-Thompson, E. Bellingshausen sea ice extent recorded in an antarctic peninsula ice core. *Journal of Geophysical Research: Atmospheres* **121**, 13–886 (2016).

[27] Macha, J. *et al.* Distinct central and eastern pacific el niño influence on antarctic surface mass balance. *Geophysical Research Letters* **51**, e2024GL109423 (2024).

[28] Vannitsem, S., Dalaiden, Q. & Goosse, H. Testing for dynamical dependence: Application to the surface mass balance over antarctica. *Geophysical Research Letters* **46**, 12125–12135 (2019).

[29] Wang, S. *et al.* Strong impact of the rare three-year la niña event on antarctic surface climate changes in 2021–2023. *npj Climate and Atmospheric Science* **8**, 173 (2025).

[30] Bodart, J. & Bingham, R. The impact of the extreme 2015–2016 el niño on the mass balance of the antarctic ice sheet. *Geophysical Research Letters* **46**, 13862–13871 (2019).

[31] Bailey, A., Singh, H. K. & Nusbaumer, J. Evaluating a moist isentropic framework for poleward moisture transport: Implications for water isotopes over antarctica. *Geophysical Research Letters* **46**, 7819–7827 (2019).

[32] Wang, H. *et al.* Influence of sea-ice anomalies on antarctic precipitation using source attribution in the community earth system model. *The Cryosphere* **14**, 429–444 (2020).

[33] Sodemann, H. & Stohl, A. Asymmetries in the moisture origin of antarctic precipitation. *Geophysical research letters* **36** (2009).

[34] Trusel, L. D., Kromer, J. D. & Datta, R. T. Atmospheric response to antarctic sea-ice reductions drives ice sheet surface mass balance increases. *Journal of Climate* **36**, 6879–6896 (2023).

[35] Hofsteenge, M., Cullen, N., Sodemann, H. & Katurji, M. Synoptic drivers and moisture sources of snowfall in coastal victoria land, antarctica. *Journal of Geophysical Research: Atmospheres* **130**, e2024JD042021 (2025).

[36] Belušić, D. *et al.* Hclim38: a flexible regional climate model applicable for different climate zones from coarse to convection-permitting scales. *Geoscientific Model Development* **13**, 1311–1333 (2020).

[37] Zwally, H. J. *et al.* Mass gains of the antarctic ice sheet exceed losses. *Journal of Glaciology* **61**, 1019–1036 (2015).

[38] Shepherd, A. *et al.* A reconciled estimate of ice-sheet mass balance. *Science* **338**, 1183–1189 (2012).

[39] Otosaka, I. N. *et al.* Mass balance of the greenland and antarctic ice sheets from 1992 to 2020. *Earth System Science Data Discussions* **2022**, 1–33 (2022).

[40] Döhne, T., Horwath, M., Groh, A. & Buchta, E. The sensitivity kernel perspective on grace mass change estimates. *Journal of Geodesy* **97**, 11 (2023).

[41] Ditmar, P. Conversion of time-varying stokes coefficients into mass anomalies at the earth's surface considering the earth's oblateness. *Journal of Geodesy* **92**, 1401–1412 (2018).

[42] Ivins, E. R. *et al.* Antarctic contribution to sea level rise observed by grace with improved gia correction. *Journal of Geophysical Research: Solid Earth* **118**, 3126–3141 (2013).

[43] Wang, X. *et al.* Comparing surface mass balance and surface temperatures from regional climate models and reanalyses to observations over the antarctic ice sheet. *International Journal of Climatology* e8767 (2025).

[44] van Dalum, C. T., van de Berg, W. J., van den Broeke, M. R. & van Tiggelen, M. The surface mass balance and near-surface climate of the antarctic ice sheet in racmo2. 4p1. *EGUsphere* **2025**, 1–40 (2025).

[45] Mo, R. Edara: An era5-based dataset for atmospheric river analysis. *Scientific Data* **11**, 900 (2024).

[46] Simmonds, I., Keay, K. & Tristram Bye, J. A. Identification and climatology of southern hemisphere mobile fronts in a modern reanalysis. *Journal of Climate* **25**, 1945–1962 (2012).

[47] Guan, B. & Waliser, D. E. Detection of atmospheric rivers: Evaluation and application of an algorithm for global studies. *Journal of Geophysical Research: Atmospheres* **120**, 12514–12535 (2015).

[48] Bénard, P. *et al.* Dynamical kernel of the aladin–nh spectral limited-area model: Revised formulation and sensitivity experiments. *Quarterly Journal of the Royal Meteorological Society: A journal of the atmospheric sciences, applied meteorology and physical oceanography* **136**, 155–169 (2010).

[49] Lenderink, G. & Holtslag, A. A. An updated length-scale formulation for turbulent mixing in clear and cloudy boundary layers. *Quarterly Journal of the Royal Meteorological Society: A journal of the atmospheric sciences, applied meteorology and physical oceanography* **130**, 3405–3427 (2004).

[50] Bengtsson, L. *et al.* The harmonie–arome model configuration in the aladin–hirlam nwp system. *Monthly Weather Review* **145**, 1919–1935 (2017).

[51] Pinty, J.-P. & Jabouille, P. A mixed-phase cloud parameterization for use in mesoscale non-hydrostatic model: simulations of a squall line and of orographic precipitations 217–220 (1998).

[52] Müller, M. *et al.* Arome-metcoop: A nordic convective-scale operational weather prediction model. *Weather and Forecasting* **32**, 609–627 (2017).

[53] De Rooy, W. C. & Siebesma, A. P. A simple parameterization for detrainment in shallow cumulus. *Monthly weather review* **136**, 560–576 (2008).

[54] Bechtold, P., Cuijpers, J., Mascart, P. & Trouilhet, P. Modeling of trade wind cumuli with a low-order turbulence model: toward a unified description of cu and sc clouds in meteorological models. *Journal of Atmospheric Sciences* **52**, 455–463 (1995).

[55] Batrak, Y., Kourzeneva, E. & Homleid, M. Implementation of a simple thermodynamic sea ice scheme, sice version 1.0-38h1, within the aladin–hirlam numerical weather prediction system version 38h1. *Geoscientific Model Development* **11**, 3347–3368 (2018).

[56] Boone, A., Calvet, J.-C. & Noilhan, J. Inclusion of a third soil layer in a land surface scheme using the force–restore method. *Journal of Applied Meteorology and Climatology* **38**, 1611–1630 (1999).